\documentclass[nofootinbib,english,aps,prd,twocolumn,showpacs,superscriptaddress,groupedaddress]{revtex4}  % for review and submission

\usepackage[T1]{fontenc}
\usepackage[latin9]{inputenc}
\usepackage{float}
\usepackage{amsmath}
\usepackage{txfonts}
\usepackage{amsfonts}
\usepackage{multirow}
\usepackage{mathrsfs}
\usepackage{subfig}
\usepackage{amsmath}
\usepackage{amssymb}
\usepackage{bm}
\usepackage{bbm}
\usepackage{xcolor}
\usepackage{babel}
\usepackage{footmisc}
\usepackage{graphicx}
\usepackage{caption}
\newcommand {\bseq}{\begin{subequations}}
\newcommand {\eseq}{\end{subequations}}
\newcommand {\pvint}[2]{{\int\!\!\!\!\!\!-}_{\!\!\!\!#1}^{#2}}
\newcommand*{\rom}[1]{\uppercase\expandafter{\romannumeral #1\relax}}
\def\Xint#1{\mathchoice
{\XXint\displaystyle\textstyle{#1}}%
{\XXint\textstyle\scriptstyle{#1}}%
{\XXint\scriptstyle\scriptscriptstyle{#1}}%
{\XXint\scriptscriptstyle\scriptscriptstyle{#1}}%
\!\int}
\def\XXint#1#2#3{{\setbox0=\hbox{$#1{#2#3}{\int}$ }
\vcenter{\hbox{$#2#3$ }}\kern-.6\wd0}}

\def\dashint{\Xint-}

\newcommand{\Langle}{\left\langle}
\newcommand{\Rangle}{\right\rangle}

\newcommand{\nn}{\nonumber}

\newcommand{\beq}{\begin{equation}}
\newcommand{\eeq}{\end{equation}}
\newcommand{\bqa}{\begin{eqnarray}}
\newcommand{\eqa}{\end{eqnarray}}

\begin{document}

\title{Gauge-invariant decomposition of meson energy in two-dimensional QCD}

\author{Yu Jia~\footnote{jiay@ihep.ac.cn}}
\affiliation{Institute of High Energy Physics, Chinese Academy of
Sciences, Beijing 100049, China}
\affiliation{School of Physics, University of Chinese Academy of Sciences,
Beijing 100049, China}

\author{Rui Yu~\footnote{yurui@ihep.ac.cn}}
\affiliation{Institute of High Energy Physics, Chinese Academy of Sciences, Beijing 100049, China}
\affiliation{School of Physics, University of Chinese Academy of Sciences,
Beijing 100049, China}

\author{Xiaonu Xiong~\footnote{x.xiong@fz-juelich.de}}
\affiliation{Institute for Advanced Simulation,
Institut f\"ur Kernphysik and J\"ulich Center for Hadron Physics,
Forschungszentrum J\"ulich, D-52425 J\"ulich, Germany}

\date{\today}

\begin{abstract}
This work is driven by our curiosity towards some basic questions in QCD:
how is the energy of a moving hadron partitioned among different gauge-invariant
sectors in QCD Hamiltonian?
How is the energy of an massless pion separated between quark and gluon sectors,
particularly in the soft pion limit?
Is it possible to decompose the celebrated Gell-Mann-Oakes-Renner (GOR) relation?
To what extent can we justify the quark potential model from the field-theoretical
mass decomposition for heavy quarkonium?
Due to limitation of contemporary nonperturbative tools,
we do not yet know answers to these questions in realistic QCD.
In this work, we take the 't Hooft model (two-dimensional QCD in large-$N$ limit) as a prototype model
that mimics some essential aspects of the true QCD.
We investigate the gauge-invariant energy decomposition of a flavor-neutral meson
that can carry an arbitrary momentum (including stationary case), with meson species ranging
from massless pion to bottomonium.
All the aforementioned questions can be addressed satisfactorily within this model, in particular
some unexpected patterns related to pion are discovered for the first time in two-dimensional QCD.
We hope that our study can offer some useful clues and stimulation for future investigations on hadron
energy decomposition in realistic QCD.
\end{abstract}

\maketitle

%-----------------------
\section{Introduction}
%-----------------------
%{\noindent\color{blue} \it Introduction.}
%-----------------------
It is of long interest to fathom the quark-gluon structure of hadrons directly from the first principle of
QCD, especially to unravel how the mass of a hadron is partitioned among the expectation values of
different sectors of the QCD Hamiltonian.
Obviously such a decomposition calls for a fully nonperturbative treatment.
The early work in this direction was pioneered by Ji~\cite{Ji:1994av,Ji:1995sv},
who carried out a gauge-invariant decomposition of nucleon
mass, with estimates based on some phenomenological inputs.
Recently, there also came out, for the first time from the angle of lattice QCD,
some exploratory studies of mass decomposition for several species of mesons,
including pion, $\rho$, and lowest-lying charmonia~\cite{Yang:2014xsa}.

In spite of a few existing studies, there remain many basic questions in ${\rm QCD}_4$
inaccessible to current nonperturbative techniques.
Such open questions include: How is the {\it energy} of a moving hadron partitioned among
different gauge-invariant sectors?
Can we have some concrete knowledge on the energy decomposition of an exactly massless pion,
particularly in the soft pion limit?
What is the microscopic decomposition of the celebrated Gell-Mann-Oakes-Renner (GOR) relation
for a pseudo Goldstone?
What is the exact correspondence between the mass decomposition for a heavy quark-antiquark
bound state and phenomenological quark potential model? The answers to all
these questions in ${\rm QCD}_4$ appear elusive in the foreseeable future.

In contrast to the notoriously intractable ${\rm QCD}_4$, two-dimensional QCD (hereafter ${\rm QCD}_{2}$)
in the large-$N$ limit, often referred to as the 't Hooft model~\cite{tHooft:1974pnl},
is a solvable toy model, which yet resembles the realistic QCD in many aspects,
such as color confinement, Regge trajectories, chiral symmetry spontaneous breaking,
quark-hadron duality, {\it etc.}.
Historically, once an interesting new feature of realistic QCD was established,
one often took 't Hooft model as a fruitful laboratory to concretise our
understanding.

In this article, we carry out a systematic investigation on the energy decomposition of a meson
in the 't Hooft model, irrespective of whether the meson is stationary or fast-moving.
We can express each gauge-invariant sector of energy decomposition
in terms of the chiral angle and forward (backward)-moving bound-state wave functions.
We can answer all the questions raised a few paragraphs above.
The most nontrivial findings of this paper
are about meson energy/mass decomposition in
the chiral limit and heavy quark limit.
It is hoped that most features observed in the 't Hooft model may carry over to the
decomposition of hadron energy in ${\rm QCD}_4$.

%-----------------------
%\section{Gauge-invariant decomposition of hadron energy}
%-----------------------
%\vspace{0.3 cm}
%-----------------------
\section{Gauge-invariant decomposition of hadron energy.}
%{\noindent\color{blue} \it Gauge-invariant decomposition of hadron energy.}
%-----------------------
The QCD lagrangian reads
%----------------------------
\beq
%----------------------------
\label{QCD:lagrangian}
%----------------------------
{\mathcal L}_{{\rm QCD}} = -\frac{1}{4} F^a_{\mu\nu} F^{a\,\mu\nu}
+\overline{\psi}\left(iD\!\!\!/ -m\right)\psi,
%----------------------------
\eeq
%----------------------------
where $\psi$ signifies the quark field, $F^a_{\mu\nu}$ is the gluon field strength tensor, and
$D_\mu=\partial_\mu-ig_s A^a_\mu T^a$ denotes the covariant derivative, with $T^a$ denoting the generators of the
$SU(N)$ group in the fundamental representation.
For simplicity, we will consider only one single flavor of quark throughout this work.

The symmetric energy-momentum tensor of QCD is then
%----------------------------
\beq
%----------------------------
T^{\mu\nu}=\frac{1}{2}\overline{\psi}i\overleftrightarrow{D}^{(\mu}\gamma^{\nu )}
\psi+\frac{1}{4}g^{\mu\nu}F^2-F^{a\,\mu\alpha}F^{a\,\nu}_{\;\;\;\;\,\alpha},
%----------------------------
\label{Belinfante:EM:tensor}
%----------------------------
\eeq
%----------------------------
where $\overleftrightarrow D = \overrightarrow D - \overleftarrow D$.
$D_\mu=\partial_\mu-ig_s A^a_\mu T^a$ denotes the color covariant derivative,
and $(\mu,\nu)$ implies the symmetrization of the Lorentz indices $\mu$ and $\nu$.

The ${\rm QCD}_2$ Hamiltonian can be broken into three pieces,
%----------------------------
\beq
%----------------------------
 H = \int\!\! dx \, T^{00} = \mathcal{M}+ \mathcal{K}+ \mathcal{G},
%----------------------------------------------
%\mathcal{T}^{00}&=\psi^{\dagger}(m\gamma^0-i\partial_1\gamma^5)\psi-\frac{1}{2}F^{01,a}F_{01}^{\ \ a}.\\
%----------------------------------------
\label{H:decompose:into:three:pieces}
%----------------------------------------
\eeq
%----------------------------
where three gauge-invariant operators are defined as
%----------------------------
\bseq
%----------------------------
\bqa
%----------------------------
&& \mathcal{M} \equiv\int \!\! dx\,  m\,\overline{\psi}\psi,
%----------------------------
\label{quark:mass:term:op}
%----------------------------
\\
%----------------------------
&& \mathcal{K} \equiv\int \!\! dx\, \frac{1}{2}\overline{\psi}i\overleftrightarrow{D}^1\gamma^1 \psi,
%----------------------------
\label{quark:kinetic:term:op}
%----------------------------
\\
%----------------------------
&& \mathcal{G} \equiv\int \!\! dx\, \left(\frac{1}{4}F^2-F^{a\,0\alpha}F^{a\,0}_{\;\;\;\;\,\alpha}\right) =
\int \!\! dx\, {1\over 2} \left(F^a_{01}\right)^2.
%----------------------------
\label{gluonic:term}
%----------------------------
\eqa
%----------------------------
\label{H:gauge:inv:decomposition}
%----------------------------
\eseq
%----------------------------
For brevity, we will often suppress the superscript $1$ in the spatial component of
two-vector $x^\mu$, {\it e.g.}, the symbols $x$ and $x^1$ will be used interchangeably,
if no confusion would arise.

We define the expectation value of an operator $\mathcal{O}$ in a mesonic momentum eigenstate $\left| P \right\rangle$
as
%----------------------------
$\Langle\mathcal O\Rangle \equiv \Langle P \left|\mathcal{O} \right|P\Rangle /\Langle P| P\Rangle$,
%----------------------------
where the meson state is relativistically normalized according to
$\Langle P|P'\Rangle=2P^02\pi\delta(P-P')$.
Sandwiching \eqref{H:decompose:into:three:pieces} between any mesonic state
carrying momentum $P$, we then expect
%----------------------------
\beq
%----------------------------
P^0 \equiv \sqrt{M^2+P^2}= \Langle\mathcal{M}\Rangle+\Langle\mathcal{K}\Rangle+\Langle\mathcal{G}\Rangle,
%----------------------------
\label{expect:value:sum:identity}
%----------------------------
\eeq
%----------------------------
with $M$ the mass of the meson.
For future convenience, we will refer $\Langle\mathcal{M}\Rangle$, $\Langle\mathcal{K}\Rangle$, and
$\Langle\mathcal{G}\Rangle$ as the quark mass term, quark kinetic energy term, and gluon energy term,
respectively.

We remark that, unlike the complication inherent in
mass decomposition for ${\rm QCD}_4$~\cite{Ji:1994av,Ji:1995sv,Yang:2014xsa},
thanks to the super-renormalizablity of ${\rm QCD}_2$, neither we need
bother to concern with the scale-dependence of the matrix elements in
\eqref{expect:value:sum:identity}, nor we need worry about the
trace anomaly associated with \eqref{Belinfante:EM:tensor}.

%-----------------------
\section{Bound-state solutions in ${\rm QCD}_{2}$}\label{sec2}
%-----------------------
%\vspace{0.3 cm}
%-----------------------
%{\noindent\color{blue} \it Bound-state solutions in ${\rm QCD}_{2}$.}
%-----------------------
Let us recapitulate some essential ingredients of the bound-state solutions of the 't Hooft model
in equal-time quantization and in axial gauge~\cite{Bars:1977ud}.
First, the large $N$ limit is specified as
%----------------------------
\beq
%----------------------------
N\to \infty,\qquad \lambda \equiv {g_s^2 N\over 4\pi}\;\;{\rm fixed}.
%----------------------------
\eeq
%----------------------------

Of central importance is the so-called mass-gap equation, which can be
obtained by diagonalizing the single-particle sector of the dressed-quark Hamiltonian,
or equivalently by minimizing the vacuum energy density~\cite{Bars:1977ud,Kalashnikova:2001df}:
%----------------------------
\beq
%----------------------------
 p\cos\theta(p)\!-\!m\sin\theta(p)\!=\! \frac{\lambda}{2}  \dashint_{-\infty}^{+\infty}\!\!dk {\sin\left[\theta(p)\!-\!\theta(k)\right]\over (p-k)^2},
%----------------------------
\label{Gap:equation}
%----------------------------
\eeq
%----------------------------
where $\dashint$ denotes the standard
principal-value prescription to sweep the infrared singularity as $k\to p$.

The integral equation \eqref{Gap:equation}
can be numerically solved to determine the chiral angle $\theta(p)$ for any given quark mass.
Note $\theta(p)$ is an odd function of $p$, and tends to $\pm \pi/2$ as $p\to \pm \infty$.
It should be stressed that the chiral angle in the massless quark limit still assumes a
nontrivial profile, which corresponds to a chiral-asymmetric vacuum
carrying a nonzero condensate.

The dressed quark then admits the following dispersive law:
%----------------------------
\beq
%----------------------------
E(p) \!=\! m\cos\theta(p)\!+\!p\sin\theta(p)\!+\!\frac{\lambda}{2} \dashint_{-\infty}^{+\infty}\!\!dk \frac{\,\cos\left[\theta(p)\!-\!\theta(k)\right]}{(p-k)^2} .
%----------------------------
\label{dispersive:law}
%----------------------------
\eeq
%----------------------------
Note for small quark mass, the dressed quark energy can even become negative for small $p$.
This is a harmless and tolerable nuisance, since the colored and gauge-variant
entity does not directly correspond to a physical quantity.

With the solutions of $\theta(p)$ and $E(p)$ available,
a pair of bound-state equations for the flavor-neutral $q\bar{q}$ mesons
were first given by Bars and Green in 1978:
%----------------------------
\bqa
%----------------------------
\label{BG:equations}
%----------------------------
&& \left[E(p)+E(P-p)\mp P_n^0)\right]\,\phi_\pm^n(p,P)
%----------------------------
\\
%----------------------------
&& =\lambda \dashint_{-\infty}^{+\infty}\!\! \frac{dk}{(p\!-\!k)^2}\!
%-----------------------------
\left[C(p,\!k,\!P)\,\phi_\pm^n(k,\!P)\!-\!S(p,\!k,\!P)\,\phi_\mp^n(k,\!P)\right].
%----------------------------
\nn
%----------------------------
\eqa
%----------------------------
where $n$ denotes the principle quantum number of the mesonic family, and $\phi^n_{\pm}$
signify the forward (backward)-moving wave functions for the $n$-th state.
The solutions of \eqref{BG:equations} determine the eigen energy of the entire mesonic family,
$P_n^0=\sqrt{M_n^2+P^2}$ for a given quark mass and meson spatial momentum $P$,
where $M_n$ denotes the mass of the $n$-th excited meson.
The trigonometric functions $C$ and $S$ in \eqref{BG:equations} are defined as~\cite{Bars:1977ud}
%----------------------------
\bseq
%----------------------------
\begin{align}
%----------------------------
& C(p,\!k,\!P)=\cos\frac{\theta(p)-\theta(k)}{2}\cos\frac{\theta\left(P-p\right)-\theta\left(P-k\right)}{2},
%----------------------------
\\
%----------------------------
& S(p,\!k,\!P)=\sin\frac{\theta(p)-\theta(k)}{2}\sin\frac{\theta\left(P-p\right)-\theta\left(P-k\right)}{2}.
%----------------------------
\end{align}
%----------------------------
\eseq
%----------------------------

Note $\dashint$ in \eqref{dispersive:law} and \eqref{BG:equations} differ from the
ordinary principle-value prescription, due to more severe IR divergences encountered there.
For a smooth test function $f(y)$,
the generalized principle-value prescription is specified as
%----------------------------
\begin{align}
%----------------------------
\notag \dashint\frac{dy}{(x-y)^2}f(y) =& \lim_{\epsilon\to 0}\Bigg[\int^{x-\epsilon} \frac{dy}{(x-y)^2}f(y)
%----------------------------
\\
%----------------------------
+&\int_{x+\epsilon} \frac{dy}{(x-y)^2}f(y)-2\frac{f(x)}{\epsilon} \Bigg].
%----------------------------
\end{align}
%----------------------------

Eq.~\eqref{BG:equations} has been numerically solved for stationary~\cite{Li:1987hx}
and moving~\citep{Jia:2017uul} mesons, for a variety of quark mass.
To simultaneously handel both stationary and moving mesons,
we deliberately choose the orthogonality condition of wave functions
that differs from \cite{Jia:2017uul,Jia:2018qee}  by a factor of $|P|$:
%----------------------------
\begin{align}
%----------------------------
&\int_{-\infty}^{+\infty}{dp}\left[\phi^+_n(p,P)\,\phi^+_m(p,P)-
\phi^-_n(p,P)\,\phi^-_m(p,P)\right]=\delta_{nm}.
%----------------------------
\label{oth}
%----------------------------
\end{align}
%----------------------------

It can be proved that~\cite{Bars:1977ud,Jia:2017uul}, with the increasing quark mass/principle quantum number/meson momentum,
the backward-moving wave functions $\phi_-$ quickly fades away.
In particular, in the $P\to\infty$ limit, \eqref{BG:equations} smoothly transition into
the celebrated 't Hooft equation~\cite{tHooft:1974pnl}:
%----------------------------
\beq
%----------------------------
 {m^2\over x(1\!-\!x)}\phi^n(x)\!-\!
2\lambda\pvint{0}{1} dy {\phi^n(y)\!-\!\phi^n(x)\over (x\!-\!y)^2} \!=\!  M_n^2 \phi^n(x),
%----------------------------
\label{thooft}
%----------------------------
\eeq
%----------------------------
where $\phi_+^n(p,P)$ in the infinite-momentum frame
can be identified with the 't Hooft light-cone wave function (LCWF)
$\phi^n(x)$, with $x\equiv p/P$ denoting the fraction of the quark momentum with
respect to the meson momentum.

%-----------------------
\section{Energy decomposition of a moving meson}\label{secax}
%-----------------------
%\vspace{0.3 cm}
%-----------------------
%{\noindent\color{blue} \it Energy decomposition of a moving meson.}
%-----------------------
A particular advantage of the axial gauge $A^{1a}=0$ is that,
the $0$-component of the gauge potential
becomes a constrained rather than a dynamical variable:
%----------------------------
\beq
%----------------------------
%----------------------------
A^{0a}(x^0,x)= {g_s\over 2}\!\!\int\!\! dy \left\vert x-y \right\vert
\psi^\dagger(x^0,y)T^a\psi(x^0,y),
%----------------------------
\label{A0a:constrained}
%----------------------------
\eeq
%----------------------------
which is nothing but the instantaneous Coulomb potential in two dimensions.
Moreover, when specializing to the axial gauge,
$\mathcal{M}$ in \eqref{quark:mass:term:op} remains intact,
the other two in \eqref{H:gauge:inv:decomposition} reduce to
%----------------------------
\bseq
%----------------------------
\bqa
%----------------------------
%----------------------------
&& \mathcal{K}= -\int \!\!dx\, \frac{1}{2}\overline{\psi}i\overleftrightarrow{\partial_1} \gamma^1\psi,
%----------------------------
\label{Quark:kinetic:axial:gauge}
%----------------------------
\\
%----------------------------
&& \mathcal{G}=\int\!\! dx \,{1\over 2} \left(\partial_1 A^0\right)^2
%----------------------------
\\
%----------------------------
&& \quad = -{g_s^2\over 4} \!\! \int \!\!\!\! \int \!\! dx\,dy\,
\psi^\dagger(x)T^a\psi(x)\,|x-y|\,\psi^\dagger(y)T^a\psi(y).
\nn
%----------------------------
\label{Gluonic:energy:axial:gauge}
%----------------------------
\eqa
%----------------------------
\eseq
%----------------------------

To compute hadronic matrix elements involving these operators,
rather than utilize the diagrammatic Bethe-Salpeter approach~\cite{Bars:1977ud},
it turns much more transparent to invoke the operator method,
that is, the bosonization technique equipped with
Bogoliubov transformation~\cite{Kalashnikova:2001df}.
In fact, these techniques
have recently been applied to systematically investigate
the partonic quasi distributions in 't Hooft model~\cite{Jia:2018qee}.
Following the steps expounded in \cite{Kalashnikova:2001df,Jia:2018qee},
and going through some tedious but straightforward
algebras, we end up with the following expectation values
when sandwiched between the $n$-th mesonic state:
%----------------------------
\bseq
%----------------------------
\begin{align}
%--------------------------------------M
& \Langle\mathcal{M}\Rangle_n\! = m\!\! \int^\infty_{-\infty}\!\! dp\big[\cos\!  \theta  (p)\!+\!\cos\!  \theta  (\bar{p})\big]
\!\left[\left(\phi^n_+(p,\!P)\right)^2\!  +\! \left(\phi_-^n(p,\!P)\right)^2\right],
%----------------------------
\label{quark:mass:energy:final}
%----------------------------
\\
%----------------------------
&\Langle\mathcal{\,K\,}\Rangle_n\! = \!\! \int^\infty_{-\infty}\!\! dp\big[p \sin\!  \theta  (p)\!+\!\bar{p} \sin\!
\theta  (\bar{p})\big]
%--------------------------------------K
%----------------------------K
\!\left[\left(\phi_+^n(p,{P})\right)^2\!  +\! \left(\phi_-^n(p,P)\right)^2\right],
%----------------------------
\label{quark:kinetic:energy:final}
%----------------------------
\\
%---------------------------------------G
& \Langle\mathcal{\,G\,}\Rangle_n\! = \! \frac{\lambda}{2}
 \int^\infty_{-\infty} \!\! dp \dashint^\infty_{-\infty} \! dk
\frac{1}{(k\!-\!p)^2}\Bigg\{ \big[\cos  (\theta (k)-\theta  (p))
\nn\\
%---------------------------------------G
&+ \cos  (\theta  (\bar{k})-\theta  (\bar{p}))\big]\left[\left(\phi_+^n(p,\!{P})\right)^2
+\left(\phi_-^n(p,\!{P})\right)^2\right]
%---------------------------------------G
\label{gluon:energy:final} \\
%---------------------------------------G
& - 2C(p,k,P)\Big[\phi_+^n(k,\!{P})\phi_+^n(p,\!{P})\!+\!\phi_-^n(k,\!{P})\phi_-^n(p,\!{P})\Big]
%---------------------------------------G
\nn\\
%---------------------------------------G
& + 2S(p,k,P)\Big[\phi_+^n(k,\!{P})\phi_-^n(p,\!{P})\!+\!\phi_-^n(k,\!{P})\phi_+^n(p,\!{P})\Big]\Bigg\},
%-----------------------------------------
\nn
%----------------------------
\end{align}
%----------------------------
\label{Energy:decomposition:master:formula}
%----------------------------
\eseq
%----------------------------
with $\bar{p}\equiv P-p$, $\bar{k} \equiv P-k$.
With the aid of \eqref{BG:equations} and orthogonality conditions,
one readily verifies that the sum of three pieces
in \eqref{Energy:decomposition:master:formula} indeed recover $P_n^0=\sqrt{M_n^2+P^2}$.

Eq.~\eqref{Energy:decomposition:master:formula} is the key formula of this paper, which provides a unified
energy decomposition formula, valid for any quark mass and any meson momentum. In the following sections,
we will explore the consequence of \eqref{Energy:decomposition:master:formula} in
several different settings.

We cannot resist to remark that, several auxiliary quantities appearing
in intermediate stages, such as \eqref{dispersive:law} and
\eqref{A0a:constrained}, are gauge-dependent and plagued with infrared divergences,
and by default we have employed the principle-value prescription as a specific
IR regulator. In contrast, each individual piece in \eqref{Energy:decomposition:master:formula}
is gauge-invariant as well as IR finite thence regulator independent,
which can thereby be endowed with some physical significance.

%-----------------------
\section{Energy decomposition in infinite-momentum limit}
%\vspace{0.3 cm}
%-----------------------
%{\noindent\color{blue} \it Energy decomposition in infinite-momentum limit.}
%-----------------------
It is curious to know how the energy is partitioned among three components
with the ever increasing meson momentum.
In general, boosting a bound-state wave function is a highly nontrivial, dynamical
rather than a kinematic operation.
Nevertheless, the energy decomposition simplifies substantially
when the meson is boosted to the infinite-momentum frame.
In the $P\to \infty$ limit, with $x=p/P$ kept fixed,
it is legitimate to replace $\phi_+^n(p,P)$ by the LCWF $\phi^n(x)$,
throw away all the $\phi_-^n$ terms, and take the
approximation $\tan \theta(x P)\to {x P\over m}+{\cal O}(1/P)$~\cite{Bars:1977ud,Jia:2017uul}.
%and $E(x P)\to |x| P+{m^2-2\lambda\over |x|P}+{\cal O}(1/P^2)$.
Eq.~\eqref{Energy:decomposition:master:formula} in the $P\to \infty$ limit then simplifies
into
%----------------------------
\bseq
%----------------------------
\begin{align}
%----------------------------
&\lim_{P\rightarrow\infty}\Langle\mathcal{M}\Rangle_n = \frac{m^2}{P}\int_0^1\!\!dx {\left({\phi^n}(x)\right)^2\over x(1-x)},
%----------------------------
\\
%----------------------------
&\lim_{P\rightarrow\infty}\Langle \mathcal{K}\Rangle_n = P-
{m^2\over 2P}\int_0^1\!\!dx\ \frac{\left({\phi^n}(x)\right)^2}{x(1-x)},
%----------------------------
\\
%----------------------------
&\lim_{P\rightarrow\infty}\Langle \mathcal{G}\Rangle_n
= \frac{\lambda}{P}\!\!\int_0^1\!\! dx\dashint_0^1dy\,\frac{\phi^n(x)\!-\!\phi^n(y)}{(x-y)^2}\phi^n(x).
%----------------------------
%----------------------------
\end{align}
%----------------------------
\label{Energy:decompe:IMF:limit}
%----------------------------
\eseq
%----------------------------

As an alternative to fathom the behavior of a meson with infinite momentum,
one can directly start from the light-cone Hamiltonian.
From~\eqref{Belinfante:EM:tensor}, one can decompose the
${\rm QCD}_2$ light-cone Hamiltonian into
%----------------------------
\beq
%----------------------------
 H_{\rm LC} = \int\!\! dx^- \, T^{+-} = \mathcal{M}_{\rm LC}+\mathcal{G}_{\rm LC},
%----------------------------------------
\label{LC:H:decompose:into:two:pieces}
%----------------------------------------
\eeq
%----------------------------
where the light-cone coordinates are defined as $x^\pm=\frac{x^0\pm x^1}{\sqrt{2}}$.
The two gauge-invariant operators are
%----------------------------
\beq
%----------------------------
\mathcal{M}_\mathrm{LC}\!=\!\int\! dx^-\frac{m}{2}\overline{\psi}\psi,\qquad
%----------------------------
\mathcal{G}_\mathrm{LC}\!=\!\frac{1}{2}\int\! dx^- F^{a\,+-}F^a_{-+}.
%----------------------------
\eeq
%----------------------------
\label{LC:H:gauge:inv:decomposition}
%----------------------------
%----------------------------
Note there is no counterpart of the quark kinetic energy \eqref{quark:kinetic:term:op},
in the light-cone decomposition.

One can compute the expectation values of these operators along a similar path which leads to
\eqref{Energy:decomposition:master:formula}.
Imposing the light-cone gauge $A^{a\,+}=0$, and employing the light-cone Hamiltonian
operator approach~\cite{Jia:2018qee}, one finally finds
%----------------------------
\bseq
%----------------------------
\bqa
%----------------------------
&\Langle\mathcal{M}_{\rm LC}\Rangle_n\!\! &= {P\over P^+} \lim_{P\rightarrow\infty} \left[\Langle\mathcal{M}\Rangle_n
+\Langle {\mathcal K}\Rangle_n -P\right],
%----------------------------
\\
%----------------------------
&\Langle\mathcal{G}_{\rm LC}\Rangle_n\!\! &= {P\over P^+}\lim_{P\rightarrow\infty} \Langle\mathcal{G}\Rangle_n,
%----------------------------
%----------------------------
\eqa
%----------------------------
\label{LC:matrix:elements:related:to:ordinary:ME}
%----------------------------
\eseq
%----------------------------

Following \eqref{expect:value:sum:identity}, one readily confirms
%----------------------------
$
%----------------------------
P^-_n = \Langle\mathcal{M}_{\rm LC}\Rangle_n + \Langle\mathcal{G}_{\rm LC}\Rangle_n,
%----------------------------
$
%----------------------------
as expected.

%----------------------------
\begin{figure}
\begin{centering}
\includegraphics[clip,width=0.51\textwidth]{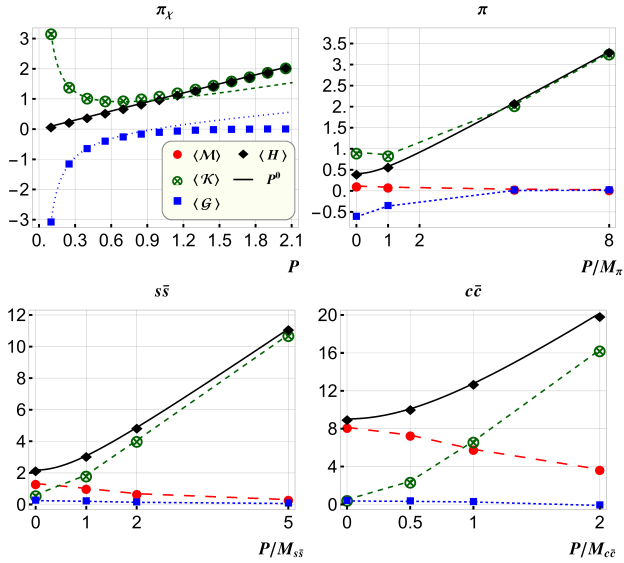}
\caption{Numerical predictions for the energy decomposition of four types of ground-state mesons,
as a function of the meson momentum. The solid curve denotes $P^0=\sqrt{P^2+M^2}$.}
\label{Fig:lowest-lying:energy:decomp:vs:momentum}
\end{centering}
\end{figure}
%----------------------------

%-----------------------
\section{Numerical results}
%-----------------------
%\vspace{0.3 cm}
%-----------------------
%{\noindent\color{blue} \it Numerical results.}
%-----------------------
For numerical calculations, we fix the 't Hooft coupling $\lambda=0.18/\pi\;{\rm GeV}^2$, in conformity to the
accepted value of string tension in the realistic ${\rm QCD}_4$~\cite{Jia:2017uul}. For simplicity,
all masses will be given in units of $\sqrt{2\lambda} =340$ MeV.
Following \cite{Jia:2017uul,Jia:2018qee},
we consider four different values of quark masses, and concentrate on
the corresponding four different lowest-lying mesons: chiral (massless) pion $\pi_\chi$, physical pion $\pi$, a fictitious strangeonium,
and charmonium. For physical pion and charmonium, quark mass is tuned such that the realistic
masses of $\pi$ and $J/\psi$ are correctly reproduced.
The numerical solutions for the chiral angle and Bars-Green equations for these
cases were described in detail in \cite{Jia:2017uul}.
It is then a straightforward exercise to plug these numerical solutions into
the integrals in \eqref{Energy:decomposition:master:formula},
to accomplish the energy decomposition.

In Fig.~\ref{Fig:lowest-lying:energy:decomp:vs:momentum}, we plot the energy decompositions against the meson momentum,
for the aforementioned four types of mesons. It is ready to recognize the pattern in the large momentum limit,
as demanded in \eqref{Energy:decompe:IMF:limit}:
as a consequence of Lorentz contraction of the meson's spatial extent,
the quark mass term $\Langle\mathcal{M}\Rangle$ and gluonic term $\Langle\,\mathcal{G}\,\Rangle$ decreases as $1/P$,
but $\Langle\,\mathcal{K}\,\Rangle$, scaling as $P$, completely saturate the meson energy.
When the meson momentum gets small, there also arise some peculiar patterns.
For heavy quarkonium, unsurprisingly, its rest mass is overwhelmed by the quark mass term, and receives positive
contribution from the gluon term.
For lighter meson such as pion, the gluonic term $\Langle\,\mathcal{G}\,\Rangle$ becomes deeply negative.

%-----------------------
\section{Energy decomposition of soft chiral pion and physical pion}
%\vspace{0.3 cm}
%-----------------------
%{\noindent\color{blue} \it Energy decomposition of chiral pion in soft limit.}
%-----------------------
Coleman's theorem states that there cannot arise Goldstone boson in two spacetime dimensions~\cite{Coleman:1973ci}.
Nevertheless, ${\rm QCD}_2$ in the large $N$ limit appears to be very peculiar, in the sense that
the ground-state meson for massless quark is massless, and the theory also admits a nonzero
chiral condensate in the chiral limit~\cite{Zhitnitsky:1985um,Li:1986gf}. Thus
it is customary to still refer the massless pseudoscalar meson as the
chiral pion ($\pi_\chi$), a would-be ``Goldstone'' particle.
The Bars-Green wave functions for the chiral pion,  $\phi^{\pi_\chi}_\pm$, are analytically
known in terms of the chiral angle~\cite{Kalashnikova:2001df}, which turn to be nonanalytic at $P=0$~\cite{Jia:2017uul}.
Note for soft pion, the backward-moving wave function is not suppressed with respect to
the forward-moving one at all, reflecting the nontrivial vacuum structure of
the 't Hooft model in the chiral limit~\cite{Kalashnikova:2001df}.

From the up-left panel of Fig.~\ref{Fig:lowest-lying:energy:decomp:vs:momentum}, it is amazing to see
that $\Langle\,\mathcal{K}\,\Rangle$ and $\Langle\,\mathcal{G}\,\Rangle$ tend to $\pm \infty$ when $P\rightarrow0$!
Fitting the numerical data from $P=10^{-4}$ to $10^{-2}$,
these two components are well parameterized by the following form:
%----------------------------
\beq
%----------------------------
\Langle \mathcal{K} \Rangle_{\pi_\chi} =  \frac{0.312}{P}+0.661 P,\;\;
%----------------------------
\Langle \mathcal{G} \Rangle_{\pi_\chi} =   -\frac{0.312}{P}+0.339 P.
%----------------------------
\eeq
%----------------------------
The divergences cancel when summing up quark kinetic and gluonic pieces.
The origin of this singularity can be traced to the huge amplitudes of $\phi_\pm^{\pi_\chi}$
in the $P\to 0$ limit,
as a consequence of almost identical profiles
of these two wave functions for soft pion.
This can be understood from the peculiar minus sign in the normalization condition
\eqref{oth}, a characteristic of
Bogoliubov transformation~\cite{Kalashnikova:2001df}.

It is certainly curious to speculate whether similar
singular behavior for massless pion also arise in ${\rm QCD}_4$ or not.
Unfortunately, it is beyond the contemporary Monte Carlo simulation technique to
directly implement the massless quark on the lattice.

%----------------------------
%\section{Mass decomposition of the stationary pseudo-Goldstone pion}
%----------------------------
%\vspace{0.3 cm}
%----------------------------
%{\noindent\color{blue} \it Mass decomposition of the ``pseudo Goldstone''.}
%----------------------------
We then turn to the mass decomposition of the physical pion, the so-called pseudo-Goldstone boson.
Intriguingly, the physical $\pi$ in 't Hooft model is found to obey the followng relation very well:
%-----------------------
\beq
%-----------------------
f_\pi^2 M_\pi^2 = - 4m \langle \Omega\vert \overline{\psi}\psi \vert\Omega \rangle,
%-----------------------
\label{GOR:relation}
%-----------------------
\eeq
%-----------------------
with $f_\pi=\sqrt{N/\pi}$, $\langle \bar{\psi}\psi \rangle =-\sqrt{2\lambda}N/\sqrt{12}$ in the chiral limit~\cite{Zhitnitsky:1985um,Li:1986gf}.
Eq.~\eqref{GOR:relation} is just the two-dimensional counterpart of the celebrated
GOR relation~\cite{GellMann:1968rz},
directly reflecting the pseudo-Goldstone nature of the pion.

\begin{figure}
\begin{centering}
\includegraphics[clip,width=0.4\textwidth]{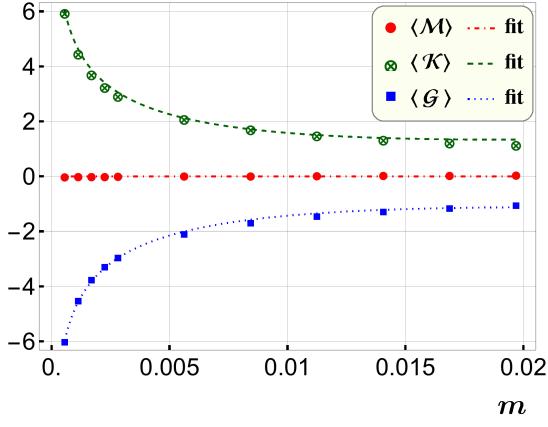}
\caption{The light quark mass dependence of $\Langle\mathcal{M}\Rangle$, $\Langle\,\mathcal{K}\,\Rangle$ and $\Langle\,\mathcal{G}\,\Rangle$, associated with the mass decomposition of
a stationary ``pseudo-Goldstone'' pion. }.
\label{Fig:pseudo:Goldstone:mass:breakup}
\end{centering}
\end{figure}

We would like to examine the microscopic origin for the GOR relation in our case.
Fig.~\ref{Fig:pseudo:Goldstone:mass:breakup} presents the light quark mass dependence of
the mass decomposition for the ``pseudo Goldstone'' boson.
In the $m\to 0$ limit, our numerical results reveal there emerge the power-law divergence
in quark kinetic term and gluonic term.
Fitting the numerical data in the small $m$ range,
the three mass components are found to be well described by
the following parameterizations~\footnote{To be more general,
we may parameterize $\Langle\mathcal{M}\Rangle_\pi $ in the form $c\sqrt{m}+\alpha m^{\beta}$. Nevertheless,
the fitting leads to $\beta=0.495\approx 1/2$, thus these two terms can
be merged into a single entity $\propto \sqrt{m}$.}:
%----------------------------
\bseq
%----------------------------
\begin{align}
%----------------------------
\Langle\mathcal{M}\Rangle_\pi = &0.448 \sqrt{m} + {\cal O}(m),
%----------------------------
\\
%----------------------------
\Langle\mathcal{K}\Rangle_\pi =&\hphantom{-\;\,}{1.005}{m^{-0.288}}+6.252 \sqrt{m}-2.719+ {\cal O}(m),
%----------------------------
\\
%----------------------------
 \Langle\mathcal{G} \Rangle_\pi =& -{1.005}{m^{-0.288}}-4.781 \sqrt{m}+2.719+ {\cal O}(m).
%----------------------------
\end{align}
%----------------------------
\label{pseudo-Goldstone:mass:decomposition}
%----------------------------
\eseq
%----------------------------
The nonanalytic $m^{-0.288}$ terms, which exhibit some weak logarithmical singularities,
cancel upon summing $\Langle\mathcal{K}\Rangle$ and $\Langle\mathcal{G}\Rangle $.
Each of the component in \eqref{pseudo-Goldstone:mass:decomposition} contains
a term $\propto \sqrt{m}$, so all of them contribute to GOR relation.
Note the sigma term $\Langle \overline{\psi}\psi \Rangle_\pi \propto 1/\sqrt{m}$, is power
divergent in the chiral limit.
Summing three pieces in \eqref{pseudo-Goldstone:mass:decomposition}, we obtain
$M_\pi=1.919 \sqrt{m}$, in decent agreement with the GOR relation $M_\pi=1.905 \sqrt{m}$,
with a relative error: 0.7\%.

It is certainly worth speculating whether the decomposition of GOR relation
in ${\rm QCD}_4$ exhibits similar pattern. 
With steady technical advance,
it seems feasible for the lattice simulation to examine the mass decomposition of
a physical pion in the real world in near future.

%-----------------------
\section{Mass decomposition of stationary heavy quarkonium}
%\vspace{0.3 cm}
%-----------------------
%{\noindent\color{blue} \it Breakup of heavy quarkonium binding energy.}
%-----------------------
For a heavy quark, since $m\gg \sqrt{2\lambda}$, one may neglect the interaction term
in \eqref{Gap:equation} and \eqref{dispersive:law}, consequently the chiral angle $\theta(p)\approx \tan^{-1}{\frac{p}{m}}$ and
$E(p)\approx \sqrt{m^2+p^2}$. In the rest frame of a heavy quarkonium,
the typical velocity of the heavy quark is quite small, so one may further make nonrelativistic expansion
for $\theta(p)$ and $E(p)$.
Setting the $\theta$ angle to 0 and approximating $E(p)\approx m+{p^2\over 2m}$,
and dropping $\phi_-$,  one can reduce the Bars-Green equations \eqref{BG:equations}
into a single equation:
%-----------------------
\beq
%-----------------------
\frac{p^2}{2\mu}\phi_+^n(p) -\lambda\pvint{-\infty}{\infty}\frac{dk}{(p-k)^2}\phi_+^n(k)=
{\cal E}_n\,\phi_+^n(p),
%-----------------------
\label{Schroedinger:eq:momentum:picture}
%-----------------------
\eeq
%-----------------------
with the reduced mass $\mu=m/2$ and binding energy ${\cal E}_n \equiv M_n-2m$.
This is nothing but the Schr\"odinger equation with a linear potential
$V(x)=\lambda\pi\left\vert x \right\vert$ in the momentum space,
with $\phi_+^n(p)$ identified with the corresponding momentum-space Schr\"odinger wave function.
The equivalent Schr\"odinger equation in the coordinate space reads
%-----------------------
\begin{align}
%-----------------------
\left(\frac{-\partial_x^2}{2\mu}+\lambda\pi\left| x \right| \right)\psi_n(x) =  \mathcal{E}_n \psi_n(x),
%-----------------------
\end{align}
%-----------------------
with $\psi_n(x)$ the coordinate-space wave function for the $n$th excited state.
The $\psi_n(x)$s are known analytically:
%-----------------------
\begin{align}
%-----------------------
\psi_n(x)=
N_n\times\begin{cases}
 \hphantom{(-1)^n}\mathrm{Ai}\left(\frac{{\mu^{1/3} } (2 \pi  \lambda  x-2 \mathcal{E}_n )}{(2 \pi \lambda)^{2/3} }\right) & x < 0,
%-----------------------
 \\
%-----------------------
 (-1)^n\mathrm{Ai}\left(\frac{{\mu^{1/3} } ( -2 \pi  \lambda  x-2 \mathcal{E}_n)}{(2 \pi \lambda)^{2/3} }\right) & x> 0,
 %-----------------------
\end{cases}
%-----------------------
\end{align}
%-----------------------
where $\mathrm{Ai}$ is the Airy function, $N_n$ is the normalization constant
to guarantee $\int\! dx\, \psi^2_n(x) =1$.
The energy spectrum of the heavy quarkonium family
can be determined through the the equations:
%-----------------------
\begin{align}
%-----------------------
\begin{cases}
%-----------------------
\mathrm{Ai}'\!\left(-\tfrac{\mathcal{E}_n\left(2\mu\right)^{1/3}}{\left(\pi\lambda\right)^{2/3}}\right)=0 &\text{even }n,
%-----------------------
\\
%-----------------------
\mathrm{Ai}\left(-\tfrac{\mathcal{E}_n\left(2\mu\right)^{1/3}}{\left(\pi\lambda\right)^{2/3}}\right)=0  &\text{odd }n.
%-----------------------
\end{cases}
%-----------------------
\end{align}
%-----------------------

\begin{figure}
%-----------------------
\begin{centering}
%-----------------------
\includegraphics[clip,width=0.47\textwidth]{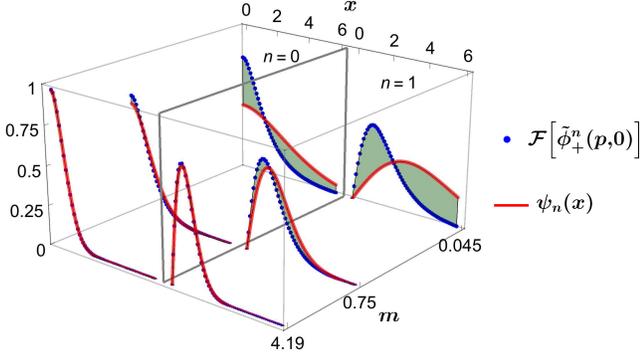}
\caption{Comparison between the Fourier-transformed $\phi^n_+\left(p,0\right)$ and
the coordinate-space Schr\"odinger wave function $\psi_n(x)$,
with three different quark mass $m=0.045$, $0.75$, $4.19$ corresponding to $u$, $s$, $c$, respectively.
Both ground state $n=0$ and the 1st excited state $n=1$ are plotted.
$\mathcal{F}$ denotes the Fourier transform introduced in \eqref{eq:QMvsQFT}.
}
\label{Fig:QMvsQFT}
%-----------------------
\end{centering}
%-----------------------
\end{figure}

Apparently, in the limit of an infinitely heavy quark,
the  Schr\"odinger wave function $\psi_n(x)$ is linked with
the forward-moving Bars-Green wave function in rest frame,
$\phi^{n}_+\left(p,0\right)$, through the Fourier transform
%-----------------------
\beq
%-----------------------
\label{eq:QMvsQFT}
%-----------------------
\psi_n(x) = \lim_{m\rightarrow\infty} \int^{\infty}_{-\infty} \!\!\frac{dp}{2\pi} \phi^{n}_+(p,0)\,e^{ipx}.
%-----------------------
\eeq
%-----------------------
We devote Fig.~\ref{Fig:QMvsQFT} to quantify the difference between $\psi_n(x)$,
obtained by solving the Schr\"odinger equation,
and the Fourier-transformed  $\phi^{n}_+(p,0)$, obtained by solving Bars-Green equations,
for a variety of quark mass. One clearly sees that, when $m$ reaches the charm quark mass,
these two wave functions converge to each other satisfactorily.

%-----------------------
%-----------------------

\begin{figure}
\begin{centering}
\includegraphics[clip,width=0.4\textwidth]{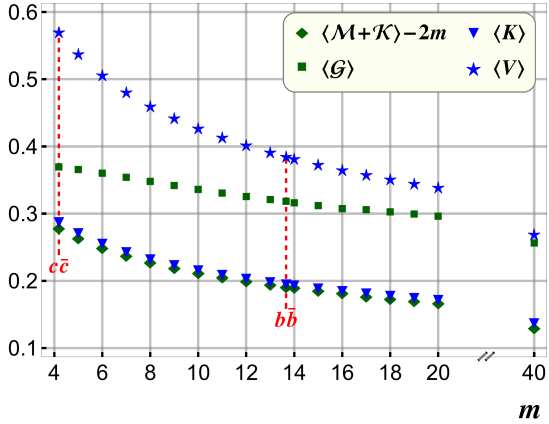}
\caption{Comparison between the kinetic and potential energy in quantum mechanics
and those field-theoretical mass components in \eqref{QFT:vs:QM:kinetic:potential:energy}.
Only the ground state quarkonium ($n=0$) is considered. }
\label{Fig:quarkonium:mass:breakup}
\end{centering}
\end{figure}

In a similar vein, expanding \eqref{Energy:decomposition:master:formula}
in the heavy quark limit, we obtain
%-----------------------
\bseq
%-----------------------
\bqa
%-----------------------
\lim_{m\rightarrow\infty} \Langle \mathcal{M}\Rangle_n \!&=&\!2m-\int_{-\infty}^{\infty}\!\! dp\, \frac{p^2}{2\mu}{\phi^n_+}(p)^2,
%-----------------------
\\
%-----------------------
\lim_{m\rightarrow\infty} \Langle \mathcal{K}\Rangle_n \!&=&\!\!\int_{-\infty}^{\infty}\!\! dp\, \frac{p^2}{\mu}\phi^n_+(p)^2,
%-----------------------
\\
%-----------------------
\lim_{m\rightarrow\infty} \Langle \mathcal{G}\Rangle_n \!&=&\!\!\lambda\!\!\int_{-\infty}^{\infty}\!\! dp\,\pvint{-\infty}{\infty}\!dk\frac{{\phi_+^n}(p)^2\!-\!\phi_+^n(k)\phi^n_+(p)}{(p-k)^2}.
%-----------------------
\eqa
%-----------------------
\label{mass:breakup:heavy:quark:limit}
%-----------------------
\eseq
%-----------------------
We thus can identify the expectation values of the kinetic and potential energy of
the $n$-th state in quantum mechanics
with the field-theoretical matrix elements in \eqref{mass:breakup:heavy:quark:limit}:
%----------------------------
\bseq
%----------------------------
\begin{align}
%----------------------------
\Langle K\Rangle_n  \equiv & \left\langle n \left| {\hat{p}^2\over 2\mu} \right|n \right\rangle
=\int\!\!dx\,\psi_n(x){-\partial_x^2\over 2\mu}\psi_n(x)
%----------------------------
\nn\\
%----------------------------
=&\lim_{m\rightarrow\infty} \left[\Langle \mathcal{M}\Rangle_n+ \Langle \mathcal{K}\Rangle_n \right]- 2m,
%----------------------------
\label{kinetic:energy:def}
%----------------------------
\\
%----------------------------
\Langle V\Rangle_n  \equiv& \Big\langle n\Big| \,\lambda\pi\vert x \vert \, \Big|n \Big\rangle =\int\!\!dx\, \psi_n(x)\lambda\pi\left|x\right|\psi_n(x)
%----------------------------
\nn \\
%----------------------------
 =& \lim_{m\rightarrow\infty} \Langle \mathcal{G}\Rangle_n,
%----------------------------
\label{potential:energy:def}
%----------------------------
\end{align}
%----------------------------
\label{QFT:vs:QM:kinetic:potential:energy}
%----------------------------
\eseq
%%----------------------------
where $ \Langle V\Rangle_n =2  \Langle K\Rangle_n$ in line with virial theorem.
From Fig.~\ref{Fig:quarkonium:mass:breakup}, one sees that the
kinetic energy obtained in single-particle quantum mechanics already
agrees well with $\Langle \mathcal{M}\Rangle+\Langle \mathcal{K}\Rangle-2m$ for $c\bar{c}$.
Nevertheless, to have a decent agreement between potential energy and
$\Langle \mathcal{G}\Rangle$,
the quark appears to be at least three times heavier than the $b$ quark.
It is interesting to note that, although the Schr\"odinger wave function and
the forward-moving Bars-Green wave function already coincide in shape for the $c\bar{c}$ family, $\left\langle\mathcal{G}\right\rangle$ and $\left\langle V \right\rangle$ exhibit much slower
convergence tendency. This may indicate that, for modestly heavy quark,
the true profile of the chiral angle may still play an indispensable
role on mass breakup of quarkonium.

%-----------------------
\section{Summary}
%-----------------------
%\vspace{0.3 cm}
%-----------------------
%{\noindent\color{blue} \it Summary.}
%-----------------------
We have studied the gauge-invariant energy decomposition of a flavor-neutral meson within the framework of 't Hooft model.
The energy of a meson can be decomposed into three gauge-invariant components, {\it i.e.},
quark mass term, quark kinetic term, and gluonic term, each of which can be expressed in terms of the
chiral angle and the forward (backward)-moving bound-state wave functions.
In the chiral limit, we find an amazing feature, that the quark kinetic energy and gluonic energy for a
massless ``Goldstone pion'' diverge as $\pm 1/P$ as the meson momentum $P\to 0$.
For small yet nonzero quark mass, we observe that both the quark kinetic and gluonic energy of
the ``pseudo-Goldstone'' pion possess nonanalytic power-law dependence of $m$ together with $\sqrt{m}$.
We are able to conduct a microscopic decomposition for the GOR relation in the two-dimensional QCD.
In the heavy quark limit, we illustrate how the field-theoretical decomposition of the binding energy for a heavy quarkonium can recover the familiar non-relativistic quantum mechanics.
It is hoped that these interesting features observed in ${\rm QCD}_2$,
can offer some inspiration and serve as a prototype, for our understanding of
energy decomposition of hadrons in realistic ${\rm QCD}_4$.

\begin{acknowledgments}
{\noindent\it Acknowledgment.}
%-----------------------
%-----------------------
The work of Y.~J. and R.~Y. is supported in part by the National Natural Science Foundation of China under Grants No.~11475188,
No.~11621131001 (CRC110 by DFG and NSFC). The work of X.-N.~X. is supported by the Deutsche Forschungsgemeinschaft (Sino-German CRC 110).
%-----------------------
\end{acknowledgments}

\end{document}